# A Fast Concurrent Power-Thermal Model for Sub-100nm Digital ICs

J.L. Roselló[†], V. Canals[†], S.A. Bota[†], A. Keshavarzi[‡] and J. Segura[†]

[†]Electronic Technology Group. Universitat de les Illes Balears. Campus UIB. 07122 Palma. Spain
[‡]Mircoprocessor Research Labs, Intel Corp., Portland (OR)
email:j.rossello@uib.es

**Abstract**

As technology scales down, the static power is expected to become a significant fraction of the total power. The exponential dependence of static power with the operating temperature makes the thermal profile estimation of high-performance ICs a key issue to compute the total power dissipated in next-generations. In this paper we present accurate and compact analytical models to estimate the static power dissipation and the temperature of operation of CMOS gates. The models are the fundamentals of a performance estimation tool in which numerical procedures are avoided for any computation to set a faster estimation and optimization. The models developed are compared to measurements and SPICE simulations for a 0.12µm technology showing excellent results.

## 1. Introduction

Power dissipation has become a major concern in IC design due to the increasing importance of portable devices and wireless communication systems, and to the heating problems that may arise in high-density/high-performance circuits. High operating temperature degrades IC performance and impacts reliability. Both power dissipation and operating frequency are worsened at high temperatures due to the increase of leakage currents and carrier mobility reduction. Moreover, circuit density and complexity may lead to spatial temperature gradients within the IC, thus impacting power differently at different IC regions.

Technology scaling rules based on constant field scaling dictate voltage supply reduction from generation to generation. Supply voltage scaling requires threshold voltage reduction to maintain the gate delay reduction, with a side effect of a dramatic increase of leakage current given its exponential dependence with threshold voltage. Projections show that in the 90nm process generation node, subthreshold leakage power can contribute as much as 42% of the total power [1]. Consequently, leakage power is no longer negligible in deep-submicron CMOS technologies and the development of models for an accurate estimation of this power component is a must.

This work has been partially supported by the Spanish Ministry of Science and Technology, the Regional European Development Funds (FEDER) from the EU project TIC2002-01238, and an Intel Laboratories-CRL research grant

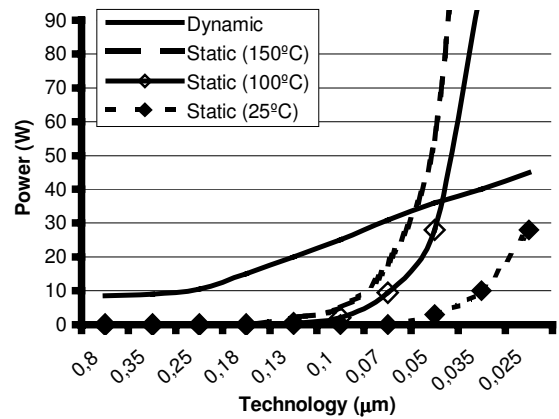

**Fig. 1** Power dependence with technology scaling at three different temperatures. Figure reproduced with permission of Dr. Duarte

Static power is exponentially dependent with the operating temperature (Fig. 1) where it is shown that static power will surpass the dynamic one implying a higher dependence of power with temperature [2]. For this reason, the development of fast electro-thermal CAD tools will be necessary for sub-100nm digital technologies since temperature will have a paramount influence on the overall power. Moreover, the power estimation and optimization of high-density ICs with hundreds of millions of transistors on a single chip requires CAD tools based on a compact modeling with analytical expressions rather than numerical approaches (as SPICE simulations) since analytical solutions provide faster estimations, thus minimizing the impact on the total design cycle.

Different works have been published during last years to derive analytically the operating temperature of MOSFETs or the static power dissipation. Sharma et al. derived in [3] a self-heating model by solving a two-dimensional heat diffusion equation. The accuracy of this model was compared to measurements from a 1.1µm pMOSFET in [4], showing an overestimation of about a 200%. Ostermeir et al. [5] developed a model for the estimation of the temperature distribution in the MOS transistor. The heat source of the channel was computed numerically as the integral of discrete line sources giving an analytical solution for only a single line source. More recently, Sabry et. al. provided a lumped thermal model for self-heating in MOSFETs [6]. The model was expressed as a function of complex surface integrals without providing closed-form expressions.

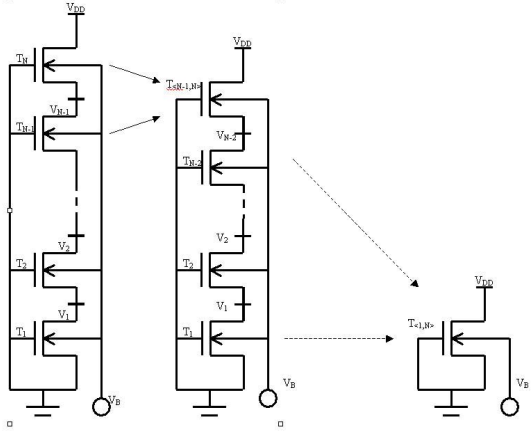

Fig. 2 Chain collapsing scheme. Each pair of transistors at the top are collapsed into a single equivalent transistor

An analytical model for the estimation of the static power dissipation was presented by Richard Gu et. al. in [7]. The model can be applied only to gates with up to three serially connected transistors and assumes that the drain-source voltage of each transistor is much greater than the thermal voltage $V_T$. A more general analytical model for the estimation of the standby leakage power of transistor stacks is presented in [8]. The model can be applied to gates with an indeterminate number of serially connected transistors and shows reasonable results with respect SPICE estimations. More recently, an analytical model for the leakage power prediction was presented in [9]. The model is only valid for gates with no more than two serially connected transistors and assumes that the drain-source voltage of each transistor is much more greater than the thermal voltage $V_T$.

In this paper we develop a compact analytical model to estimate the operating temperature and the static power dissipated by CMOS gates. These models can be combined to analytical models for the dynamic power [10] to estimate analytically the thermal profile and the total power. The rest of the paper is organized as follows: in section 2 we develop the static power model, section 3 presents a fast analytical method to estimate the thermal profile of ICs. Finally, we present the results and the conclusions in sections 4 and 5 respectively.

## 2. Power estimation of CMOS gates

The power dissipated in CMOS gates can be divided in two main components: the dynamic and the static power dissipation. The dynamic is associated to the switching activity of each gate and two power components can be differentiated:

- **Transient power**. Due to the charge and discharge of the effective output capacitance. Usually, the transient power is computed as $P_t = \alpha f C V_{DD}^2$, where $\alpha$ is the switching activity factor, f the frequency of operation, C the output capacitance, and $V_{DD}$ the supply voltage.

- **Short-circuit power.** This component is due to the direct current path between the supply and the ground node present during the gate input transition. It depends on the input transition time and on several process parameters. It can be accurately estimated using the model from [10].

The static power dissipation is usually neglected with respect dynamic power as long as $V_{th} \gg V_T$. The progressive reduction of the threshold voltage makes this component more significant for each technology generation. An accurate analytical model for the estimation of the static component is presented in the next subsection.

### 2.1 Compact model for static power estimation

*1) General definitions*

We assume that the main static power source is due to subthreshold currents. For one single transistor the subthreshold current is given by [11]:

$$I_{OFF} = \frac{W}{L} I_0 \left(\frac{T}{T_{ref}}\right)^2 e^{\frac{V_{GS}-V_{TH}}{nV_T}} \left(1 - e^{-\frac{V_{DS}}{V_T}}\right) \quad (1)$$

where W and L are the channel width and length respectively, T is the temperature of operation, $T_{ref}$ is a reference temperature, $I_0$ is a process-dependent parameter, while $V_T$, $V_{GS}$, $V_{DS}$, and $V_{TH}$ are the thermal, gate-source, drain-source and threshold voltage respectively. The threshold voltage may be expressed as:

$$V_{TH} = V_{T0} + \gamma' V_{SB} + K_T (T - T_{ref}) - \sigma(V_{DS} - V_{DD}) \quad (2)$$

where $V_{T0}$ is the zero bias threshold voltage, $\gamma'$ is related to the body effect, $K_T$ is the sensibility of the threshold voltage with temperature, while $\sigma$ accounts for the DIBL effect.

The determination of the static current through the whole CMOS gate requires a computation of the current through each branch of transistors connecting the supply and ground nodes.

We define an OFF branch as a chain of serially connected transistors with at least one transistor being in the OFF state. An ON branch is defined as a chain of serially connected transistors where all devices are ON. For each OFF chain we find an equivalent transistor with an effective width such that its OFF current (given by (1)) is equal to the current through the whole OFF chain.

If an OFF chain is in parallel with an ON chain then it is discarded for the static current estimation. Each OFF chain is collapsed to a single equivalent transistor (the collapsing technique is detailed in the next subsection) with an effective chain width. Finally, two OFF chains connected in parallel are collapsed into a single equivalent transistor with an effective width equal to the sum of the effective widths of the equivalent transistor of each OFF chain.

*2) Gate collapsing technique*

In Fig. 2 we show an OFF chain of 'N' nMOS transistors (for an OFF chain of pMOS transistors the analysis is equivalent). In the analysis we consider only the OFF transistors while the ON transistors are considered to be part of the internal nodes of the chain. The closer to ground transistor and the upper transistor are labeled as $T_1$ and $T_N$ respectively, while the internal nodes are labeled from $V_1$ to $V_{N-1}$. The upper transistor is connected to the supply voltage $V_{DD}$ while the substrate is assumed to have a voltage $V_B$.

The transistor collapsing method is applied as follows: the pair of transistors at the top of the chain are collapsed into a single equivalent transistor $T_{<N-1,N>}$ leading to a chain with N-1 OFF transistors. The process is repeated until we obtain a single transistor $T_{<1,N>}$ with an equivalent width $W_{<1,N>}$ such that its current is equal to the OFF current of the original chain. In the next subsection we explain in detail the collapsing process of two serially-connected transistors.

*3) Collapsing two series-connected transistors*

To determinate the width of the equivalent transistor $T_{<N-1,N>}$ we evaluate the current through $T_N$ and $T_{N-1}$. Following (1) and (2) the current through $T_N$ (defined as $I_N$) is given by:

$$I_N = \frac{W_N}{L} I_0 \left(\frac{T}{T_{ref}}\right)^2 e^{\frac{-V_{T0}+\gamma' V_B-(1+\sigma+\gamma')V_{N-1}}{nV_T}} \quad (3)$$

where the exponential factor dependent on $V_{DS}/V_T$ can be neglected as long as $V_{DD} \gg V_T$. The current through $T_{N-1}$ would be:

$$I_{N-1} = \frac{W_{N-1}}{L} I_0 \left(\frac{T}{T_{ref}}\right)^2 e^{\frac{-V_{T0}+\gamma' V_B-(1+\sigma+\gamma')V_{N-2}+\sigma(V_{N-1}-V_{DD})}{nV_T}} \left(1 - e^{\frac{V_{N-1}-V_{N-2}}{V_T}}\right) \quad (4)$$

The equivalent transistor $T_{<N-1,N>}$ will have a current expression similar to (3) since its drain-source voltage will be much larger than $V_T$.

$$I_{<N-1,N>} = \frac{W_{<N-1,N>}}{L} I_0 \left(\frac{T}{T_{ref}}\right)^2 e^{\frac{-V_{T0}+\gamma' V_B-(1+\sigma+\gamma')V_{N-2}}{nV_T}} \quad (5)$$

From (3) we obtain $W_{<N-1,N>}$ as:

$$W_{<N-1,N>} = W_N e^{\frac{-(1+\sigma+\gamma')(V_{N-1}-V_{N-2})}{nV_T}} \quad (6)$$

Therefore, as stated by (6), the effective width of the two transistors is exponentially dependent on $V_{N-1}-V_{N-2}$. For the estimation of $V_{N-1}-V_{N-2}$ we equate expressions (3) and (4) and solve for $V_{N-1}$. Unfortunately this problem has no analytical solution, although for some cases we can obtain analytical approximations:

a) $V_{N-1}-V_{N-2} \gg V_T$. For this case we obtain the next expression:
$$V_{N-1} - V_{N-2} \cong \alpha V_T f(W_N, W_{N-1}) \equiv V_A \quad (7)$$

b) $V_{N-1}-V_{N-2} < V_T$. For this second case we have:
$$V_{N-1} - V_{N-2} \cong V_T e^{f(W_N, W_{N-1})} \equiv V_B \quad (8)$$

where $f(W_N, W_{N-1})$ and $\alpha$ take the form:

$$f(W_N, W_{N-1}) = Ln\left(\frac{W_N}{W_{N-1}}\right) + \frac{\sigma V_{DD}}{nV_T} \quad (9)$$

$$\alpha = \frac{n}{1+\gamma'+2\sigma}$$

An empirical solution that includes both cases is given by:

$$V_{DS}^{T_{N-1}} \equiv V_{N-1} - V_{N-2} = V_T \left\{1 + \frac{(\alpha-1)e^f}{\alpha-1+e^f}\right\} Ln(1+e^f) \quad (10)$$

In Fig. 3 we show the fitness of the proposed expression with respect to the exact solution for a two transistors stack using a 0.12μm technology.

The drain-source voltage of any other transistor $T_i$ is obtained similarly as in (10). Then, the effective channel width of the equivalent transistor of the OFF chain can be obtained from (3) as:

$$W_{<1,N>} = W_N e^{\frac{-(1+\sigma+\gamma')V_{N-1}}{nV_T}} \quad (11)$$

Where $V_{N-1}$ is obtained as:

$$V_{N-1} = \sum_{i=1}^{N-1} V_{DS}^{T_i} \quad (12)$$

For a given input vector to the gate (say vector 'i') an effective width $W^i_{eff}$ is obtained using the collapsing technique described. Then the $I_{OFF}$ current of the gate is given by:

$$I_{OFF} = \frac{W^i_{eff}}{L} I_0 \left(\frac{T}{T_{ref}}\right)^2 e^{\frac{-V_{T0}-K_T(T-T_{ref})+\gamma' V_B}{nV_T}} \quad (13)$$

## 3. Thermal profile estimation of ICs

The estimation of the operating temperature for sub-100nm designs is required to get an accurate computation of the total power dissipated by the circuit. The estimation of the thermal profile in the IC can be obtained at different levels of abstraction depending on the required granularity. At the lower level the temperature is estimated for each transistor in the circuit that is used as an elementary heat source. At a higher level of abstraction an entire circuit block can be considered as a heat source. In general, the study is focused to get the thermal profile generated from the heat dissipated inside a square of dimensions $W \times L$.

The thermal distribution over the surface of the substrate can be estimated by solving the three-dimensional heat diffusion equation in the steady state:

$$\nabla \bullet \mathbf{q} = s \quad (14)$$

where $\mathbf{q}$ is the heat flux and $s$ is the power dissipated per unit volume. The heat flux $\mathbf{q}$ is proportional to the temperature gradient as:

$$\mathbf{q} = -\nabla kT \quad (15)$$

where $k$ is the thermal conductivity, and $T$ is the temperature. Expressions (14) and (15) can be used to estimate the thermal profile of the IC.

In Fig. 4 we show a square of dimensions $W \times L$ (that could represent a transistor or a large logic block) dissipating a power P. The heat generated in the square is transferred through the substrate toward the heat sink at the bottom of the IC. The solution of (15) depends on the IC boundary conditions. We assume that the heat flux orthogonal to the sides and the top of the IC is null (see Fig. 4) while the temperature at the bottom of the IC is constant.

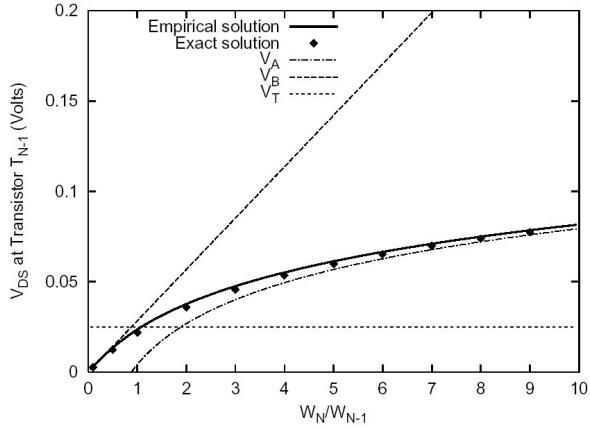

**Fig. 3 Drain-source voltage at transistor $T_{N-1}$. Expression (10) is found to be a good approximation to $V_{N-1}-V_{N-2}$**

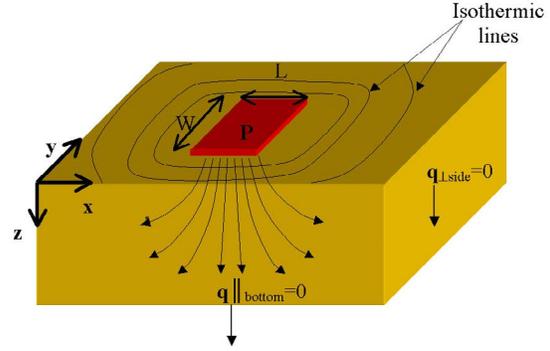

**Fig. 4 Heat diffusion scheme of a square dissipating a power P**

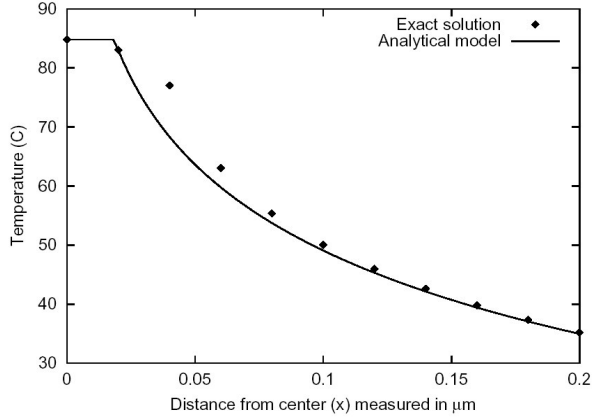

**Fig. 5 Comparison between the exact thermal profile and the approximated one for a single MOS transistor**

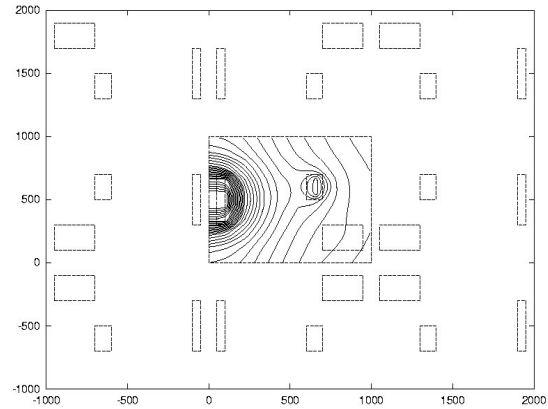

**Fig. 6 Heat sources used to compute the thermal profile of an IC with three logic blocks**

### 3.1 Temperature estimation at the middle of the square

The temperature distribution through the silicon created by a ideal punctual power source of value *P* located at the surface of the substrate is easily obtained taking into account that the heat flux orthogonal to the top of the IC is null. Solving (14) and (15), with $s(r \neq 0)=0$ and $s(r=0)=P$ we have:

$$T(\mathbf{r}) = \frac{P}{2\pi k_{Si} \|\mathbf{r}\|} \quad (16)$$

For the case of a square we assume that the power P is dissipated uniformly through this area. Then, the temperature distribution through the substrate is estimated from:

$$T(x,y) = \frac{P}{WL} \int_{-W/2}^{W/2} \int_{-L/2}^{L/2} \frac{dx_0 dy_0}{2\pi k_{Si}\sqrt{(x-x_0)^2 + (y-y_0)^2}} \quad (17)$$

Expression (17) cannot be solved analytically but an exact solution is obtained at the middle of the square (*x=y=0*).

$$T(r=0) = T_0 = \frac{P}{2\pi k_{Si} WL} \left\{ Ln\left[\frac{L+\sqrt{L^2+W^2}}{-L+\sqrt{L^2+W^2}}\right] W + Ln\left[\frac{W+\sqrt{L^2+W^2}}{-W+\sqrt{L^2+W^2}}\right] L \right\} \quad (18)$$

Equation (18) is a first-order approximation to the temperature of operation of a square with dimension *W×L* that is dissipating a power *P*.

### 3.2 Thermal profile far away from the square

Although equation (17) has no analytical solution, an approximated expression can be obtained for distances far away from the square *(|r|>> W, L)*. Under this assumption the square can be treated as a single power source and expression (16) can be used. Nevertheless we can obtain a more precise expression if we assume that *W > L*.

$$T_{line}(x,y) = \frac{P}{2\pi k_{Si} W} Ln\left[\frac{\sqrt{4x^2+W^2-4yW+4y^2}+W-2y}{\sqrt{4x^2+W^2+4yW+4y^2}-W-2y}\right] \quad (19)$$

Equation (19) is found to provide very good results (also in those cases in which *W=L*).

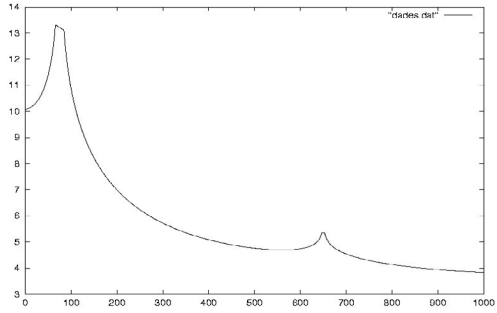
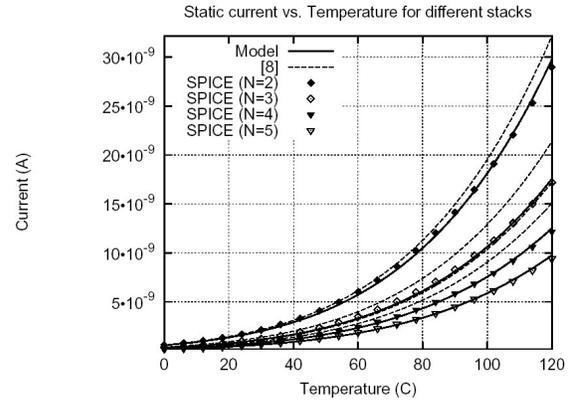

Fig. 7 Temperature distribution at the middle of the IC. The derivative of the temperature (and therefore the heat flux) at the two sides of the IC is zero.

Fig. 8 The proposed model and the model in [8] are compared with SPICE simulations for four stacks of nMOS transistors

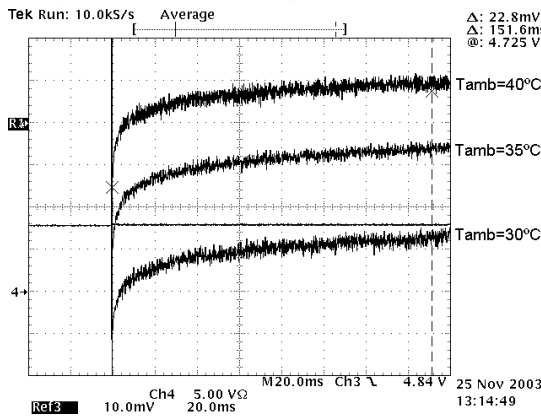
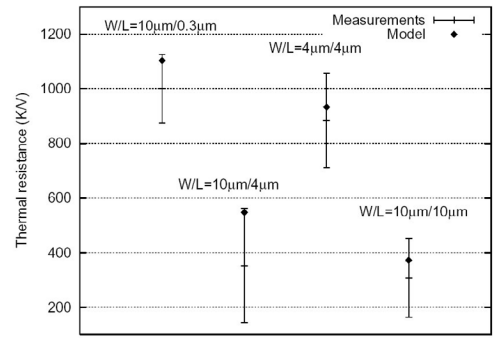

Fig. 9 Self heating measurements of a single MOS transistor at three different ambient temperatures

Fig. 10 Comparison between model predictions (dots) and self-heating measurements (bars) for four different transistors

### 3.3 General expression for the thermal profile

As $|r|$ approaches zero, equation (19) diverges. For this value of $|r|$ the temperature saturates to the value provided by equation (18). Therefore, an analytical approximation for the solution of (17) is obtained combining the solutions provided by (18) and (19).

$$T(x, y) = Min\{T_0, T_{line}(x, y)\} \quad (20)$$

A comparison between (20) and the numerical solution of (17) is plotted in Fig. 5. The thermal profile corresponds to a transistor with dimensions $W=1\mu m$ and $L=0.1\mu m$ that is dissipating $10mW$. The accuracy obtained is enough for the estimation of the thermal profile for large ICs.

For M rectangles with dimensions $W_i \times L_i$ located at $(x_i, y_i)$ and dissipating a power $P_i$ we can estimate the temperature at each point of the substrate surface using superposition:

$$T(x, y) = \sum_{i=1}^{M} Min\{T_0^i, T_{line}(x - x_i, y - y_i)\} \quad (21)$$

The solution provided by (21) is obtained assuming that the substrate is semi-infinite and the heat flux toward the top of the IC is zero (the assumption of zero heat flux through the sides of the IC and constant temperature at the bottom of the IC is still not considered). The boundary conditions are taken into account using the method of images. Consider two identical power sources with value P separated a given distance d. The heat flux across the surface located between the two power sources at d/2 is zero. To take into account the boundary conditions in an IC we use several images for each side of the IC.

This is illustrated in Fig. 6 where three different logic blocks are located within an IC with dimensions $1mm \times 1mm$. The images of each block are allocated symmetrically with respect each side of the IC. Different isothermal lines are represented in Fig. 6 showing that the heat flux **q** (orthogonal to the isothermal lines) is tangent at each side of the IC. A cross-section of the temperature distribution at the middle of the IC illustrates that at both sides of the IC the derivative of the temperature (and therefore the heat flux) is zero (see Fig. 7).

The boundary condition setting that the heat flux is orthogonal to the bottom of the IC (see Fig. 4) is also considered using the method of images. For each block (dissipating a power P) we place an image block located symmetrically with respect the bottom of the IC (and dissipating a power –P). With these power sinks we force the heat flux at the bottom of the IC to be orthogonal.

## 4. Results

### 4.1 Static power estimation

We estimate the static current driven by a stack of N nMOS transistors using the analytical model presented in section 2. The model is compared to SPICE simulations and the model presented in [8] using a 0.12µm CMOS technology in Fig. 8. Results demonstrate that the model provides an excellent agreement with respect to SPICE, and provides better results that previous works.

### 4.2 Thermal profile estimation

We measured the thermal resistances of different nMOS transistors fabricated using a 0.35µm process. The thermal resistance can be expressed to be the relationship between the power dissipated and the self-heating temperature increment ($R_{th}= \Delta T_{S-H}/P$). The self-heating temperature increment ($\Delta T_{S-H}$) is measured through the current variation of the transistor due to self-heating (linearly dependent with temperature for small temperature changes).

In Fig. 9 we show a self-heating measurement of a single transistor. The transistor is sequentially turned ON and OFF with a frequency of 3Hz. The voltage drop at a resistance in series with the transistor (directly proportional to the drain current and therefore to temperature) is measured on an oscilloscope. In Fig. 9 we show the measurements at three different ambient temperatures (*T=30ºC, 35ºC* and *40ºC*) for temperature calibration. Measurements show an exponential increment of the device operating temperature associated to the charging process of the thermal capacitance of the transistor. From these measurements we estimate the thermal resistances for four different nMOS transistors (see Fig. 10). As can be appreciated a good agreement is obtained between measurements and the proposed thermal model.

## 5. Conclusions

For an accurate power estimation of sub-100nm circuits it will be vital to solve thermal and leakage estimation models simultaneously. Since traditional SPICE computations are two slow for ULSI circuits and transistor self-heating is not considered by SPICE, compact analytical models for electro-thermal simulation of ULSI circuits is a must. In this paper we propose and validate a compact modeling solution for electro-thermal estimation of the leakage power dissipated by CMOS logic circuits. The models can be combined with analytical models of dynamic power [10] to analytically estimate the thermal profile and the total power dissipated by ULSI circuits.